\documentclass[conference]{IEEEtran}
\IEEEoverridecommandlockouts
\usepackage{cite}
\usepackage{amsmath,amssymb,amsfonts}
\usepackage{algorithmic}
\usepackage{graphicx}
\usepackage{textcomp}
\usepackage{xcolor}
\def\BibTeX{{\rm B\kern-.05em{\sc i\kern-.025em b}\kern-.08em
    T\kern-.1667em\lower.7ex\hbox{E}\kern-.125emX}}

\usepackage{fancybox}
\usepackage{amsmath, amsfonts, amsthm}
\usepackage{algorithmic}
\usepackage{soul}
\usepackage{xcolor}
\usepackage{multirow}
\usepackage{comment}
\usepackage{listings}
\usepackage{xspace}
\usepackage{hyperref}
\usepackage{enumitem}
\usepackage{threeparttable}

\definecolor{codegreen}{rgb}{0,0.6,0}
\definecolor{codegray}{rgb}{0.5,0.5,0.5}
\definecolor{codepurple}{rgb}{0.58,0,0.82}
\definecolor{backcolour}{rgb}{0.95,0.95,0.92}
\lstdefinestyle{mystyle}{
    backgroundcolor=\color{backcolour},
    morecomment=[l]{//},
    commentstyle=\color{codegreen},
    keywordstyle=\color{magenta},
    numberstyle=\tiny\color{codegray},
    stringstyle=\color{codepurple},
    basicstyle=\ttfamily\footnotesize,
    breakatwhitespace=false,
    breaklines=true,
    captionpos=b,
    keepspaces=true,
    numbers=left,
    numbersep=5pt,
    showspaces=false,
    showstringspaces=false,
    showtabs=false,
    tabsize=2
}

\newcommand{\sw}[1]{\textcolor{purple}{{\it [Shaowei says: #1]}}}

\newcommand{\peter}[1]{\textcolor{red}{{\it [Peter says: #1]}}}

\newcommand{\ashraf}[1]{\textcolor{blue}{{\it [ashraf says: #1]}}}

\newcommand{\earlystop}{\emph{early test termination}\xspace}
\newcommand{\nostop}{\emph{no\_stop}\xspace}

\newcommand{\testearlystop}{Test$_{early}$\xspace}

\newcommand{\original}{\textit{Original}\xspace}
\newcommand{\trycatch}{\textit{Trycatch}\xspace}
\newcommand{\slicing}{\textit{Slicing}\xspace}

\newcommand{\rqone}{RQ1: How prevalent is early test termination?
}
\newcommand{\rqtwo}{RQ2: How does early test termination affect test coverage?}

\newcommand{\rqthree}{RQ3: How does early test termination affect the performance of spectrum-based fault localization?}

\newcommand{\rqfour}{RQ4: Can we develop an approach to eliminate the effect of \earlystop and improve over \trycatch?}
\usepackage{tcolorbox}

\newcommand{\rqboxc}[1]{\begin{tcolorbox}[left=2pt,right=2pt,top=2pt,bottom=2pt,colback=gray!4,colframe=gray!40!black,before skip=6pt,after skip=7pt]#1\end{tcolorbox}}

\makeatletter
\let\TPT@hookin\@gobble
\let\TPT@hookarg\@gobble
\makeatother
\begin{document}

\title{Studying the Impact of Early Test Termination Due to Assertion Failure on Code Coverage and Spectrum-based Fault Localization\\
}

\author{
\IEEEauthorblockN{1\textsuperscript{st} Md. Ashraf Uddin}
\IEEEauthorblockA{
    \textit{University of Manitoba} \\
    Winnipeg, Canada \\
    uddinma1@myumanitoba.ca
}
\and
\IEEEauthorblockN{2\textsuperscript{nd} Shaowei Wang}
\IEEEauthorblockA{
    \textit{University of Manitoba} \\
    Winnipeg, Canada \\
    shaowei.wang@umanitoba.ca
}
\and
\IEEEauthorblockN{3\textsuperscript{rd} An Ran Chen}
\IEEEauthorblockA{
    \textit{University of Alberta} \\
    Edmonton, Canada \\
    anran6@ualberta.ca
}
\and
\IEEEauthorblockN{4\textsuperscript{th} Tse-Hsun (Peter) Chen}
\IEEEauthorblockA{
    \textit{Concordia University} \\
    Montreal, Canada \\
    peterc@encs.concordia.ca
}
\and
\IEEEauthorblockN{5\textsuperscript{th} Muhammad Asaduzzaman}
\IEEEauthorblockA{
    \textit{University of Windsor} \\
    Windsor, Canada \\
    masaduzz@uwindsor.ca
}
}

\maketitle

\pagenumbering{gobble} 
\pagenumbering{arabic}
  \pagestyle{plain}

\begin{abstract}
An assertion is commonly used to validate the expected program's behavior (e.g., if the returned value of a method equals an expected value) in software testing. Although it is a recommended practice to use only one assertion in a single test to avoid code smells (e.g., Assertion Roulette), it is common to have multiple assertions in a single test. One issue with tests that have multiple assertions is that when the test fails at an early assertion (not the last one), the test will terminate at that point, and the remaining testing code will not be executed. This, in turn, can potentially reduce the code coverage and the performance of techniques that rely on code coverage information (e.g., spectrum-based fault localization). We refer to such a scenario as \earlystop. Understanding the impact of \earlystop on test coverage is important for software testing and debugging, particularly for the techniques that rely on coverage information obtained from the testing. We conducted the first empirical study on early test termination due to assertion failure (i.e., \earlystop) by investigating 207 versions of 6 open-source projects. We found that a non-negligible portion of the failed tests (19.1\%) is early terminated due to assertion failure.
Our findings indicate that \earlystop harms both code coverage and the effectiveness of spectrum-based fault localization. For instance, after eliminating \earlystop, the line/branch coverage is improved in 55\% of the studied versions, and improves the performance of two popular SBFL techniques Ochiai and Tarantula by 15.1\% and 10.7\% compared to the original setting (without eliminating \earlystop) in terms of MFR, respectively.



\end{abstract}


\section{Introduction}

Testing is an effective and practical way of finding bugs and improving software quality.
However, due to the scale of modern software and the complexity of tests, it can be difficult for developers to know the cause of test failures and propose fixes.
Prior studies have proposed various automated testing techniques that leverage code coverage and test execution information to assist developers with test failure diagnosis or repair. For example, researchers have proposed using code coverage information in failed/passed tests for tasks such as spectrum-based fault localization (SBFL)~\cite{chen2022t,widyasari2022real,wang2011search,chen2022useful,jones2002visualization,abreu2009spectrum,le2015information,wong2016survey} and automated program repair~\cite{le2016history, wen2017empirical, lin2017quixbugs, yi2018correlation, liu2021critical}. For instance, Ochiai~\cite{abreu2006evaluation} and Tarantula~\cite{tarantula}, SBFL techniques, leverage both passed/failed tests together with their coverage information to identify the faulty location in the program. In modern continuous integration pipelines, SBFL is deployed and triggered automatically when a test fails~\cite{hassan2023uniloc,kochhar2016practitioners}. Therefore, when assertion failure happens, SBFL provides additional information that helps developers narrow the search space and identify the faulty code effectively. 

In testing, assertions are commonly used to verify the program's expected behavior (e.g., if the returned value of a method equals an expected value). Although it is a recommended practice to use only one assertion in a test to avoid code smell (e.g., Assertion Roulette~\cite{AssertionRoulette}), prior studies~\cite{bavota2012empirical,kim2020empirical} found that many tests may still contain multiple assertions.
One issue with tests that have more than one assertion is that when the test fails at an early assertion, the test will terminate at that point and skip the remaining test code. This, in turn, can potentially reduce the code coverage, hindering the effectiveness of SBFL or automated program repair techniques. We refer to the scenario where a test has multiple assertions and fails at an early assertion (i.e., not the last assertion) 
as \earlystop. Understanding the impact of \earlystop on test coverage is important for software testing and debugging, particularly for the techniques that rely on coverage information obtained from the testing.


Therefore, we first conduct an empirical study to understand the prevalence of \earlystop and examine its impact on the test coverage and effectiveness of SBFL techniques (i.e., Ochiai and Tarantula) that rely on coverage information. We investigated 207 versions collected from six open-source projects used in Defect4J~\cite{defects4j} and T-Evos~\cite{chen2022t}. We formulate the following research questions in our research:

\begin{itemize}[leftmargin=6pt,topsep=2pt]
\item {\bf \rqone} \\
We investigate how prevalent \earlystop is in failed tests and the amount of test code that is skipped due to \earlystop.

\textbf{Results:} Among the failed tests, a non-negligible portion (19.1\%) is early terminated due to assertion failure, and they occur in 29\% of studied versions. Due to \earlystop, on average 15.3\% to 60.5\% of the test code is skipped.

\item{\bf \rqtwo}\\
We apply two approaches \trycatch and \slicing to enforce the continuous execution of tests even if \earlystop happens.
In \trycatch, we modify a test by adding a try-catch block surrounding each assertion in the test to enforce it to keep executing.
In \slicing, we use program slicing to disassemble a test with multiple assertions into multiple independent sub-tests, which contain only one assertion and its dependent code.
We refer to the original setting as \original. We compare the testing coverage before (\original) and after applying \trycatch and \slicing.

\textbf{Results:} In general, \earlystop negatively impacts the coverage in the subject under test in terms of all granularities (i.e., instruction, line, branch, and method). \trycatch and \slicing improve the line/branch coverage in 55\% of the studied versions compared with the \original.

\item{\bf \rqthree}\\
To evaluate the impact of \earlystop on SBFL, we select two popular techniques Ochiai and Tarantula as the representatives of SBFL techniques. We compare the performance of Ochiai and Tarantula before (\original) and after applying \trycatch and \slicing, to investigate the impact of \earlystop on SBFL.

\textbf{Results:} \earlystop negatively impacts the effectiveness of SBFL. Both \slicing and \trycatch improve the performance of SBFL by eliminating \earlystop. \slicing outperforms \trycatch for SBFL. On average, \slicing improves the performance of Ochiai and Tarantula by 15.1\% and 10.7\% compared with \original in terms of Mean First Rank (MFR), respectively.
\end{itemize}

We summarize our contribution as follows:
\begin{itemize}
    \item We conducted the first empirical study on early test termination due to assertion failure (i.e., \earlystop). We provide evidence of the prevalence of \earlystop and its negative impact on testing coverage and the effectiveness of SBFL.

    \item We evaluate using two approaches (\trycatch and \slicing) to mitigate the impact of early test failures on test coverage and demonstrate that \slicing performs better.

    \item We provide actionable suggestions to prevent \earlystop, and approaches to mitigate \earlystop if it already exists in their project.

    \item We make our dataset publicly available~\cite{DataRepository} to encourage future research on this direction.
\end{itemize}

\section{Background}\label{sec:backgrou}
In this section, we first discuss the phenomenon of early test termination and its potential impact on SBFL. Then, we provide a motivating example to illustrate the impact of early test termination.

\subsection{Early Test Termination}
To assist developers with diagnosing the causes of test failures, researchers have proposed various techniques such as SBFL~\cite{chen2022t,widyasari2022real,wang2011search,chen2022useful,abreu2009spectrum,le2015information,wong2016survey}.
However, when a test terminates prematurely, the generated test coverage and test failure messages become incomplete, which affects how developers and techniques such as SBFL locate the root cause of failures.

We first formally define early test termination. Suppose there is a test $t$ that has $n$ statements \{$s_1$, $s_2$, ..., $s_i$,..., $s_n$\}. We define $t$ as an early terminated test if $t$ terminates at a statement $s_i$, where $i < n$. A test may terminate early due to various reasons, such as errors in object initialization, method invocation, or assertion.
In this paper, we focus the study on early test termination due to assertion failure.
Although having multiple assertions in a test can be considered a test smell~\cite{AssertionRoulette}, we observed that this is a common practice in the studied systems (as shown in Table~\ref{tab:dataset}, 34.6\% to 73.3\% of the tests have multiple assertions in our studied projects). We also observe that early test termination due to assertion failure happens frequently (see more details in Section~\ref{sec:rq1}). 
In JUnit, if an assertion fails, the test will terminate immediately at that point, skipping the code that follows, and it will report the test as failed.
For instance, in Figure~\ref{lst:example1}, the test {\sf testRootEndpoints} has multiple assertions, and it stops at the third assertion. As a result, the remaining test code after the failed assertion would not be executed, which potentially leads to incomplete coverage of the subject under test. Our goal is to study the impact of early test termination due to assertion failure on test coverage and the performance of SBFL. 
For simplicity, we refer to early test failure due to assertion failure as \earlystop and the test where \earlystop happens as \testearlystop in the rest of the paper.


\subsection{Spectrum-based Fault Localization}
In this work, we focus on Spectrum-Based Fault Localization (SBFL) as the subject technique to understand the impact of \earlystop. We choose SBFL because it relies directly on the quality and completeness of the collected code coverage, and it is an essential step for automated program repair. SBFL uses statistical formulas to measure the suspiciousness of program units based on the program execution traces. Execution traces, also called program spectra, contain details of the covered statements in failed and passed tests. The suspiciousness scores computed from program spectra are then used to generate a ranked list of program units that are most likely to be responsible for the fault. A number of SBFL techniques have been developed over the decades, such as Tarantula~\cite{tarantula}, Ochiai~\cite{abreu2006evaluation}, DStar~\cite{wong2016survey}, and BARINEL~\cite{abreu2009spectrum}.
In particular, Ochiai and Tarantula are commonly used SBFL techniques because of their high effectiveness~\cite{widyasari2022real,lo2010comprehensive}.
For instance, Ochiai is calculated as:
$Ochiai(s) = \frac{e_f(s)}{\sqrt{(e_f(s) + n_f(s))(e_f(s)+e_p(s))}}$. The suspiciousness score of a statement is based on the execution of passed and failed tests. Therefore, the performance of SBFL is potentially impacted by \earlystop. For instance, suppose a test has two assertions $a_1$ and $a_2$ to test two branches $b_1$ and $b_2$ in the subject under test, respectively. If the test fails at $a_1$, then only $b_1$ is covered. If no other tests cover $b_2$, the Ochiai score for all the statements in $b_2$ will be 0.
However, $a_2$ may also fail if the test were able to continue executing, and we would miss the fault in $b_2$ because of \earlystop. 




\begin{figure}
    \begin{lstlisting}[basicstyle=\scriptsize, language=Java]
public void testRootEndpoints() throws Exception {
    UnivariateRealFunction f = new SinFunction();
    UnivariateRealSolver solver = new BrentSolver();
    // endpoint is root
    double result = solver.solve(f, Math.PI, 4);
    assertEquals(Math.PI, result, solver.getAbsoluteAccuracy());
    result = solver.solve(f, 3, Math.PI);
    assertEquals(Math.PI, result, solver.getAbsoluteAccuracy());
    result = solver.solve(f, Math.PI, 4, 3.5);
    // assertion failure, terminate here
    -> assertEquals(Math.PI, result, solver.getAbsoluteAccuracy());
    result = solver.solve(f, 3, Math.PI, 3.07);
    // assertion failure
    assertEquals(Math.PI, result, solver.getAbsoluteAccuracy());
} \end{lstlisting}
    \caption{A motivating example of early test termination occurs when an assertion fails on line 13, causing all the remaining tests and assertions to be skipped.}
    \label{lst:example1}
    \vspace{-0.1in}
\end{figure}

\subsection{A Motivating Example}\label{sec:motivatingexample}
Prior studies have leveraged code coverage to identify faulty locations, and code coverage is critical for the effectiveness of prior techniques~\cite{chen2022t,widyasari2022real,wang2011search,jones2002visualization,abreu2009spectrum,le2015information,wong2016survey}. However, \earlystop can potentially reduce the code coverage and therefore negatively impact the performance of the downstream techniques.
Let us illustrate our insight through a concrete example
shown in Figure~\ref{lst:example1}, which is a test adapted from the Apache Commons Math library\footnote{https://commons.apache.org/proper/commons-math/}, to test the root-finding functionality of the $BrentSolver$ class. The test evaluates the first assertion statement at line 6 but aborts at line 11 since the third assertion fails. The aborted execution omits the unexecuted assertions (i.e., assertions at line 14) that are in the same test.

Nevertheless, the remaining code contains additional information that may be useful for fault localization. First, a bug's root cause can be directly reflected in the code coverage. Identifying additional coverage can help differentiate suspicious program elements, which can help locate the root cause (i.e., buggy statements). In this example, if the remaining test code were executed, one additional branch would be covered, providing potentially valuable information for diagnosing the buggy statements. Moreover, executing additional assertions can also improve fault localization accuracy, since assertions verify the correctness of the program behavior~\cite{xuan2014test}.
However, the failure of earlier assertions prevents the execution of later assertions.
If the remaining assertions could be executed in Figure~\ref{lst:example1} and more assertion failures are identified, we would be able to identify additional incorrect program behavior that may help in diagnosing bugs without re-executing the tests. 
Therefore, in this study, we investigate the impact of reduced coverage on the effectiveness of coverage-based fault localization techniques (i.e., SBFL).

\section{Experimental Design}\label{sec:design}
In this section, we first present our research questions (RQs). We then present the data collection process and our approach to answering the research questions.

\subsection{Research Questions}

The goal of our study is twofold: 1) understand the prevalence of \earlystop; 2) study the impact of \earlystop on the test coverage and performance of debugging techniques that rely on test coverage (SBFL in this study). Therefore, we formulate our research to answer the following research questions:

\noindent{\bf \rqone}
In RQ1, we investigate how prevalent \earlystop is in failed tests and the amount of test code that is skipped due to \earlystop. 
In addition, we also investigate the prevalence of tests with multiple assertions, since such tests provide the soil for \earlystop. Note that, if a test only has one assertion, it usually could not be the case of \earlystop, since the assertion usually is the last line of the code (we have not observed any cases in our dataset where the last line is not an assertion).


\noindent{\bf \rqtwo}
In RQ2, we aim to investigate the impact of \earlystop on the test coverage of subject under test. More specifically, we compare the test coverage in two settings: \original setting where we do not apply any approach to eliminate the \earlystop; and \nostop setting, where we enforce \testearlystop to continue executing the rest of the code after the failed assertion(s). We introduce how we enforce \testearlystop to continue executing in Section~\ref{sec:methodology}. By understanding this, we could provide insights to practitioners into the impact of \earlystop on test coverage.



\noindent{\bf \rqthree}
In RQ3, we aim to investigate the impact of \earlystop on the performance of techniques that rely on test coverage information by comparing two settings \original and \nostop similar to RQ2. In this study, we select to study SBFL as our studied case. Note that our methodology could also be used for other techniques that rely on test coverage. By understanding this, we could provide insights to practitioners into the impact of \earlystop on SBFL. 


\subsection{Data Preparation}

\begin{table*}
\scriptsize
  \centering
    \caption{The overview statistics of our studied projects, which include two parts: the statistics of all versions of studied projects (\textit{\textbf{All}}) and those of versions where \earlystop happens (\textbf{\textit{Early test termination}})*.}
    \label{tab:dataset}
    \begin{threeparttable}
    \resizebox{\textwidth}{!}{
        \begin{tabular}{l|llll|lllll}
        \hline
        & \multicolumn{4}{c|}{\textbf{\textit{\textbf{All}}}} & \multicolumn{5}{c}{\textbf{\textit{Early test termination}}}                                 \\
        \hline
        \textbf{Project} & \textbf{Version} & \textbf{LOC} & \textbf{Avg test} & \textbf{Avg T$_{multi}$} & \textbf{Version} & \textbf{T$_{total}$} & \textbf{T$_{early}$} & \textbf{T$_{earlyAssert}$} & \textbf{C$_{noexecuted}$} \\
        \hline
        fastjson         & 12                 & 6K           & 4,874.0                  & 1691.6 (34.7\%)              & 4                  & 82                     & 32                     & 8     & 8 (46.0\%)                         \\
        jackson-core     & 9                  & 18K          & 396.6                  & 160.9 (40.6\%)                & 2                  & 31                     & 29                     & 2                            &   4.5 (15.3\%)                       \\
        jfreechart       & 3                  & 92K          & 2,178.3               & 1,472.3 (67.6\%)              & 2                  & 4                      & 3                      & 3                            &     19 (60.5\%)                      \\
        math             & 104                & 199.2K       & 2,926.6               & 1,075.9 (36.8\%)             & 29                 & 170                    & 81                     & 37                           &    20.13 (44.7\%)                     \\
        joda-time        & 26                 & 26K          & 3,926.3               & 2,673.3 (68.1\%)               & 5                  & 74                     & 32                     & 9                            &  18.2 (47.1\%)                       \\
        lang             & 53                 & 18.68K       & 1,927.0                & 1,413.1 (73.3\%)               & 18                 & 110                    & 60                     & 31                           &    13.72 (49.6\%)                       \\
        \hline
        \textbf{total}   & 207                & N/A          & N/A                    & N/A                   & 60                 & 471                    & 237                    & 90                           & N/A                      \\
        \hline
        \end{tabular}
    }
\begin{tablenotes}\footnotesize
    \item[*] For \textit{\textbf{All}}, we present the number of versions, the average number of LOC, and the average number of tests in each version, the average number and ratio of tests with multiple assertions (T$_{multi}$) for each project. For \textbf{\textit{Early test termination}}, we present the number of versions, total failed tests (T$_{total}$), early terminated tests including the ones earlier terminated due to other reasons (T$_{early}$), early terminated tests due to assertion failure (T$_{earlyAssert}$), and the average amount of code (lines of code and ratio) in a test that is not executed due to assertion failure (C$_{noexecuted}$) for each project.
\end{tablenotes}
\end{threeparttable}
\end{table*}
For our experiment, we select six open-source projects from benchmark datasets Defects4J~\cite{defects4j} and T-Evos~\cite{chen2022t}. Table~\ref{tab:dataset} gives an overview of the descriptive statistics of the projects. Defects4J contains real bugs extracted from large open-source projects, which enables controlled experiments in software debugging and testing research~\cite{defects4j}. Defects4J has been widely used for evaluating fault localization techniques in the research community~\cite{pearson2017evaluating, shamshiri2015automatically, xuan2016nopol, martinez2016astor}. We selected five projects (i.e., jackson, jfreechart, commons-math (math for short), joda-time, and commons-lang (lang for short)) from Defects4J, because they are from different domains and were used as the benchmark to evaluate various SBFL in prior studies~\cite{just2014mutants,chen2022useful}. To increase the diversity of our study, we further expanded our dataset with T-Evos, following an existing study~\cite{chen2022useful}. T-Evos provides continuous test execution and failure data to help understand testing practices and evaluate automated testing techniques. We added one more project fastjson from T-Evos to increase the dataset diversity. In T-Evos, fastjson is the one with the highest number of test cases per commit and with the highest lines of test codes. This allows us to study early test termination on projects that have rigorous and mature testing. As some of the projects from T-Evos, such as jackson-core and jfreechart, overlap with Defects4J, we studied these projects using the versions provided by Defects4J.
For each subject project, we execute the program on the selected versions provided by the benchmarks. We also verify that the number of tests is sufficient, the program builds successfully, and there is at least one failed test. In total, we collected 207 versions (i.e., commits with failed tests) from six projects for our experiment. Each project has an average number of tests ranging from 396.6 to 4,874. We summarize the basic statistics of our studied projects in Table~\ref{tab:dataset} (\textbf{\textit{All}} section).

\subsection{Methodology}\label{sec:methodology}

\begin{figure}
    \centering
    \scalebox{0.48}{ \includegraphics[width=\textwidth]{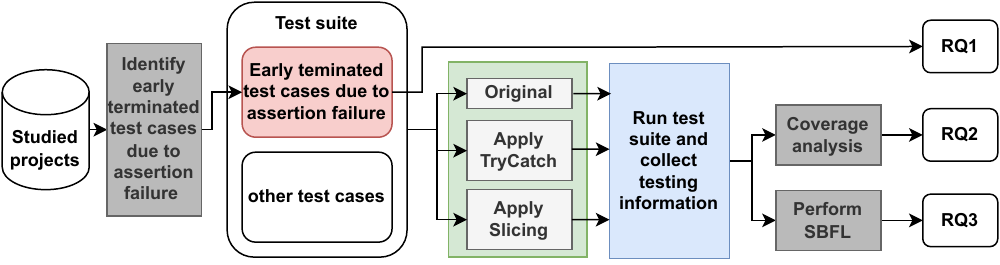}}
    \caption{The overview framework for answering the RQs.}
    \label{fig:methodology}
\end{figure}

Figure~\ref{fig:methodology} presents the overall framework for answering our RQs. First, we identify early terminated tests due to assertion failure (i.e., \testearlystop) In RQ1, we perform analysis on those identified \testearlystop. In RQ2 and RQ3, we only focus on the versions of the projects that contain \testearlystop.
In RQ2, we first run the test cases to collect testing information for the \original setting. Then, we apply \trycatch and \slicing to eliminate the \earlystop and enforce the continuous execution of \testearlystop. Next, we rerun the test, in which the \trycatch and \slicing have been applied. We compare the code coverage and the performance of SBFL of \original, \trycatch, and \slicing to answer RQ2 and RQ3. 

\subsubsection{Identifying early terminated tests due to assertion failure}\label{sec:identifyearlystop}
The first step of our study is to identify \testearlystop and their corresponding versions. For this purpose, we need to build and run all the tests on every version of our studied projects. Then, we extract and analyze the failed tests for each version. Among the failed tests, we identify the ones that are \testearlystop by examining the reports generated by JUnit. More specifically, we analyze the testing report to examine: 1) if the terminated line is in the middle of the corresponding test; and 2) if the test failure is due to assertion failure based on our definition (Section~\ref{sec:backgrou}). To this end, we collect the log information printed during testing and extract the location of each failure, meanwhile examining whether the failure is due to assertion failure by matching keywords (e.g., ``AssertionFailedError''). We identify the location of each failed assertion by analyzing the line number printed in the stack trace. 
Eventually, we identified 90 \testearlystop distributed in 60 versions (see Table~\ref{tab:dataset}). Our analysis of the RQs is based on those 60 versions that contain at least one \testearlystop.

\subsubsection{Enforcing the execution for early terminated test (\nostop setting)}\label{sec:enforcenostop}
To understand the impact of \earlystop, we need to compare two settings, where we eliminate the \earlystop (\nostop) and without eliminating \earlystop (\original). Now we introduce two approaches for \nostop and elaborate on them below.

\begin{figure}
    \begin{lstlisting}[language=Java]
@Test
public void testRootEndpoints() throws MathException {
    ...
    result = solver.solve(f, Math.PI, 4, 3.5);
    // add try-catch block
    try{
        //assertion failure
        assertEquals(Math.PI, result, solver.getAbsoluteAccuracy());
    }catch(AssertionFailedError t){
        er.add(t)
    }
    ...
    // ensure the testing result to be the same as original and re-throw the first exception
    if(er.size() > 0){
        throw er.get(0);
    }
}   \end{lstlisting}
    \caption{An example of test after applying \trycatch on the motivating example.}
    \label{lst:exampletrycatch}
\end{figure}

\noindent\textbf{\trycatch.}
We modify a test to enforce it to continous executing even if an assertion failure happens. There are several ways to achieve this goal. For instance, JUnit 5 provides ``assertAll()'' which asserts that all provided assertions get executed, even if some assertions fail. After test execution, all failures will be aggregated and reported in a MultipleFailuresError~\cite{assertall}. JUnit 4 provides ``ErrorCollector'' which allows the execution of a test to continue after the first problem is found~\cite{errorcollector}. However, ``ErrorCollector'' only supports JUnit 4 or above, and ``assertAll()'' only supports JUnit 5. Some versions of our studied projects use JUnit 3. To find a way that works for all JUnit versions, we use a try-catch block to catch the AssertionFailedError thrown by assertion failure and enforce the test to continue executing (we call this approach \trycatch). More specifically, we identify every assertion in a test by using regular expressions, such as ``assertEquals(.*)'' and ``assertTrue(.*)'', and add a try-catch block around the assertions. For instance, after applying \trycatch, the code in Figure~\ref{lst:example1} is modified as shown in Figure~\ref{lst:exampletrycatch}. We add try-catch blocks for all assertions. For space limitation, we only show the try-catch block for line 8.
Note that our modification should not change the test result (i.e., failed or passed) of the original test. For this purpose, we add an ArrayList $er$ to collect the AssertionFailedError. At the end of the test execution, if the size of the $er$ is larger than 0, which indicates at least one failure happened, we re-throw the first assertion failure in the condition block (lines 14 - 16).


\noindent\textbf{Disassembling \testearlystop using Slicing.}
Intuitively, \earlystop is caused by the dependency of statements in the test, i.e., the execution of a later assertion depends on the success of the earlier assertion. A simple approach such as \trycatch has the limitation that the results of all assertions are aggregated into one single final test result even though different assertions in the same test have different results (e.g., the final testing results become failed as long as one assertion failed, even the rest of assertion are successful), which does not provide accurate testing results and may negatively affect the SBFL. In addition, an erroneous state from an early assertion may propagate through the rest of the code and infect the state of later assertions, which makes it even harder to accurately identify the location of faulty statements in \trycatch. Prior studies show that slicing could provide a finer-grained spectrum for SBFL~\cite{alves2011fault,wen2012software,xuan2014test}. Therefore, we follow a previous study~\cite{xuan2014test} to disassemble a test with multiple assertions into multiple independent sub-tests, in which only one assertion and its dependent code are contained using program slicing~\cite{weiser1984program} (we call this approach \slicing). Compared with \trycatch, \slicing enables the independent execution of every assertion and its associated code, therefore providing finer-grained granularity than \trycatch, especially when multiple assertion failures happen in a single test and may provide more information for SBFL.

Given a test $t = \{p_1, a_1, p_2, a_2, ..., p_i, a_i, ..., p_n, a_n\}$, where $n$ is the number of assertions in $t$, $a_i$ is the $i^{th}$ assertion, and $p_i$ is the code block between two assertions $a_i$ and $a_{i-1}$, we disassemble $t$ into a set of $n$ sub-tests $T_{slice}$, i.e., $T_{slice} = \bigcup\limits_{i=1}^n slicing(t, a_i)$, where  $slicing(t, a_i)$ returns the slice of a given code $t$ from a starting point $a_i$. We replace the original $t$ with $T_{slice}$ when running tests.
Note that it is possible that there are dependencies between assertion statements. Therefore, one generated slice $slicing(t, a_i)$ could possibly include the previous assertion statements prior to $a_i$. For example, two assertions $a = assertEquals(v1.setValue(4),1)$ and $b = assertEquals(v1.getValue(),1)$. The variable $v1$ in $b$ is dependent on the method invocation $v1.setValue(4)$ in $a$. For such cases, we remove the assertion statement itself $assertEquals$, while keeping the dependent expression (s) $v1.setValue(4)$ in the assertion $b$ when performing slicing. In theory, \slicing can separate the test logic and test result, providing a more fine-grained spectrum than \trycatch.

To illustrate our approach, in Figure~\ref{lst:exampleslicing}, sliced tests are generated from the code (shown in Figure~\ref{lst:example1}) using the line ``assertEquals(Math.PI, result, solver.getAbsoluteAccuracy());''. 
We use an open-source tool JavaSlicer~\cite{JavaSlicer} to perform program slicing. More specially, given a \testearlystop, we first identify the location of each assertion statement of the \testearlystop. Then we feed the location as the starting point for slicing and the \testearlystop to JavaSlicer. JavaSlicer generates slices and we replace the given \testearlystop with the generated slices.

A threat to our study is that the program transformation possibly change the behavior of test cases. To mitigate this threat, we measured the quality of transformed tests, we applied \trycatch and \slicing on passed test cases with multiple assertions and measured their code coverage. Our results show that \original, \trycatch, and \slicing have almost the same code coverage (see more details in Section~\ref{sec:threats}).

\subsubsection{Re-running test suite and collecting testing information}\label{sec:coveragecollection}
For the RQs, we need to rerun all the tests for which we have applied \trycatch, and collect the execution information (i.e., coverage information and test results) for further analysis. Although Defect4J provides a command to get coverage information, it only provides the coverage ratio for the entire project. Our study requires statement-level coverage information for performing SBFL. Therefore, we use Jacoco~\cite{Jacoco} to collect the coverage information for each individual test. Jacoco is one of the most popular code coverage tools that instruments bytecode to trace the execution during test runs.
Jacoco has been widely used for coverage analysis in various studies~\cite{chen2022t,silva2021flacoco,saha2018bugs}, which can provide coverage information in different granularities, e.g., line, branch, method, and class. However, applying Jacoco is not straightforward and cannot be fully automated due to compilation issues, e.g., some libraries are no longer available in the central Maven repository and need to be manually downloaded, or some versions in the original setting are not compatible with Jacoco, which requires a manual update. Hence, we need to manually configure the maven build script to add the Jacoco dependency to resolve the issues and update our automation scripts accordingly. 
In our study, we conducted experiments on 207 versions of six well-known open-source projects. As one of the first studies on the impact of \earlystop, we believe our collected dataset provides valuable insight for future research.
We also made the data publicly online to encourage future studies in this direction~\cite{DataRepository}.

\begin{figure}
    \begin{lstlisting}[language=Java]
@Test
public void testRootEndpoints_1() throws Exception {
    UnivariateRealFunction f = new SinFunction();
    UnivariateRealSolver solver = new BrentSolver();
    double result = solver.solve(f, Math.PI, 4);
    assertEquals(Math.PI, result, solver.getAbsoluteAccuracy());
}
@Test
public void testRootEndpoints_2() throws Exception {
    UnivariateRealFunction f = new SinFunction();
    UnivariateRealSolver solver = new BrentSolver();
     result = solver.solve(f, 3, Math.PI);
    assertEquals(Math.PI, result, solver.getAbsoluteAccuracy());
}\end{lstlisting}
    \caption{Example of sliced tests based on the motivating example. Due to space limitation, we only present two of them.}
    \label{lst:exampleslicing}
\end{figure}

\subsection{Approaches for Answering the RQs}

\noindent{\bf Approach to answer RQ1.}
To examine the prevalence of \earlystop, we count the number of \testearlystop among failed tests. We also analyze the characteristics of \testearlystop, such as the location of the assertion failure and the amount of test code that is skipped due to \earlystop. We identify each assertion in a test and the location of each assertion failure as described in Section~\ref{sec:identifyearlystop}.

\noindent{\bf Approach to answer RQ2.}
We collect the coverage information in three settings as described in Section~\ref{sec:coveragecollection}: \original, \trycatch, and \slicing. We compare their coverage of subject under test in terms of different granularities (i.e., branches, lines, methods, and instructions). We also count the number of versions for each project, in which the coverage gets improved after applying \trycatch and \slicing in different coverage granularities.

\noindent{\bf Approach to answer RQ3.}
In this RQ, we select Ochiai and Tarantula as the SBFL techniques for evaluation, since these two techniques have good performance and are commonly used approaches for benchmarking SBFL techniques~\cite{widyasari2022real,pearson2017evaluating,zou2019empirical}. We compare the performance of both Ochiai and Tarantula in three settings: \original, \trycatch, and \slicing. We use the implementation of Ochiai and Tarantula from Flacoco~\cite{flacoco2021}.
We selected Flacoco since it supports both JUnit 3, 4, and 5. In addition, Flacoco is based on Jacoco, which aligns with the tool we use in the experiment setting. 

In this RQ, we consider three evaluation metrics: mean EXAM score (\textit{EXAM} for short)~\cite{wong2008crosstab}, Top@k~\cite{le2015constrained}, and Mean First Rank (MFR)~\cite{lou2020can}. These metrics are widely used in prior fault localization studies~\cite{widyasari2022real,pearson2017evaluating}.

\noindent\textbf{\textit{EXAM}} shows the percentage of statements that need to be inspected until reaching faulty statements. The \textit{EXAM} score is calculated as
$\frac{1}{n}\sum^n_i \frac{\textit{Rank of the faulty statement}s_i} {\textit{Total number of statement}}$,
where $n$ is the number of faulty statement, $s_i$ is the $ith$ faulty statement.
\textit{EXAM} score ranges from 0 to 1 (inclusive), with a smaller score indicating the better performance of an SBFL technique. As an illustration of the \textit{EXAM} score computation, consider the following suspiciousness score from five code statements (\(s_1\),  \(s_2\), \(s_3\), \(s_4\), and \(s_5\)), which are 0.6, 0.7, 1.0, 0.5, and 0.4. Assuming that \(s_2\) is the faulty statement that ranks second in the list, the \textit{EXAM} score will be \(\frac{2}{5} = 40\%\).  

\noindent\textbf{\em Top@k} measures the proportion of bugs with at least one faulty statement localized in the top k positions of the ranked list. We choose to include k=5 and k=10 by following prior studies~\cite{le2015constrained,kochhar2016practitioners}. A larger Top@k value indicates better performance of an SBFL technique.
Note that in the scenario when some statements have the same suspiciousness score, we assign the average rank to these statements by following prior studies~\cite{pearson2017evaluating,widyasari2022real}. The average rank is calculated as $(\frac{n}{2})+(k-1)$, where $n$ is the number of statements that have the same suspiciousness score, and k is the best rank of the statement.

\noindent\textbf{\em MFR} computes, for each project, the mean of the first relevant faulty statement rank for all bugs since the localization of the first buggy element for each bug can be crucial for localizing all faulty statements~\cite{kochhar2016practitioners,lou2020can}. A smaller MFR indicates better performance of an SBFL technique.

Similar to RQ2, we also count the number of versions for each project, in which the performance of SBFL gets improved after applying \trycatch and \slicing in terms of MFR.


\noindent{\bf Experiment Environment.} We collected our data and ran our experiments on a Linux server with one Nvidia RTX 3090 GPU, AMD Ryzen 9 12-Core CPU with 64 GB RAM, and 18TB hard drive.

\section{Results of the research questions}\label{sec:results}

\subsection*{\rqone}\label{sec:rq1}

\textbf{A significant portion of test cases have multiple assertions in a single test, ranging from 34.7\% to 73.3\% in our different studied projects.}
From Table~\ref{tab:dataset}, we observe that fastjson has the lowest percentage of tests containing multiple assertions in a single test, at 34.7\% on average.
In addition, in the two largest projects math and lang, 36.8\% and 73.3\% of the tests have multiple assertions. Our findings are compatible with prior studies~\cite{zamprogno2022dynamic,bavota2012empirical}, which show that a test having multiple assertions is prevalent in real-world projects and sets the stage for \earlystop to occur.

\noindent\textbf{Among the failed tests, a non-negligible portion (19.1\%) is due to assertion failures, occurring in 29\% of the studied versions.}
Table~\ref{tab:dataset} shows the results of early test termination in the studied projects.
Our analysis, conducted across 207 studied versions, reveals that 29\% of the versions contain at least \testearlystop, indicating that \earlystop is prevalent in the studied projects. Among the 471 failed tests observed, a significant portion (i.e., 50.3\%) was terminated early, with a non-negligible portion (i.e., 19.1\%) of these early terminated tests occurring due to assertion failure.
To better understand the reasons for early termination that are not due to assertion failure, we manually investigated 20 samples, all of which failed due to errors in object construction or method invocation.
For instance, in jfreechart, an error happens frequently when the method $clone$ of $CategoryTableXYDataset$ is invoked, causing the test to terminate early. Future studies may investigate the effect of such failures on testing practices.

\noindent\textbf{On average, 15.3\% to 60.5\% of the test code is not executed in \testearlystop due to \earlystop.}
We examine the amount of test code that is not executed due to assertion failure. From Table~\ref{tab:dataset}, we see that on average, 15.3\% to 60.5\% of code in \testearlystop is not executed due to assertion failure in our studied projects.
The results indicate that around half of the test code in \testearlystop is skipped due to \earlystop. For instance, in Figure~\ref{lst:example1}, there are in total 10 lines of statements, and the assertion fails at line 13, leaving 2 lines of code unexecuted, which accounts for 20\% of the test code.

\rqboxc{Among the failed tests, a non-negligible portion (19.1\%) is due to assertion failures and it occurs in 29\% of the studied versions. On average, 15.3\% to 60.5\% of the test code is not executed in \testearlystop due to \earlystop in our studied projects.}

\subsection*{\rqtwo}\label{sec:rq2}


\textbf{In general, \earlystop negatively impacts the coverage in the production code in terms of all granularities (i.e., instruction, line, branch, and method).} Table~\ref{tab:RQ2_comparisonValueFailedTests} presents the production code coverage by only \testearlystop of \original, \trycatch, and \slicing. We observe significant coverage improvement after eliminating \earlystop. On average, applying \trycatch and \slicing improves the production code coverage by 4.5\%, 5.3\%, 7.8\%, and 5.9\% over \original in terms of instruction, line, branch, and method, respectively. For instance, in jfreechart, after eliminating \earlystop, 16 (15.1\%) more lines, 4.5 (25.5\%) more branches, and 5 (17.5\%) more methods are covered compared to \original, in which 106 lines, 17 branches, and 28.5 methods are covered. This is reasonable, since in jfreechart, an average of 60.5\% of the test code is skipped due to \earlystop. Intuitively, if there are fewer tests in the test suite and assertions fail earlier in the testing process, the impact of \earlystop is likely to be even more pronounced.

\begin{table}
    \caption{Comparison of the production code coverage by only the \testearlystop in the settings of \original, \trycatch, and \slicing.}

    \label{tab:RQ2_comparisonValueFailedTests}
    \footnotesize
    \begin{threeparttable}
    \resizebox{0.5\textwidth}{!}{
        \begin{tabular}{l|rrrr}
        \hline
        \multicolumn{5}{c}{\textbf{\original}}                                                          \\
        \hline
        \textbf{Project} & \textbf{Instruction} & \textbf{Line}   & \textbf{Branch} & \textbf{Method} \\
        \hline
        math             & 5,521
             & 333.97
         & 127.28
         & 47.59
          \\
        joda-time        & 10,117.20

                & 2,450.80

          & 767.40

          & 568.80

             \\
        jfreechart       & 388.50

                & 106

           & 17
         & 28.50

            \\
        jackson          & 3,961

                & 854

            & 276.50
            & 140

           \\
        fastjson         & 10,614

                & 1,815

            & 604.75

            & 220.25

             \\
        lang             & 467.78

                 & 65.28

             & 24.44

           & 12.06
           \\
        \hline
        \multicolumn{5}{c}{\textbf{\trycatch and \slicing (improvement)}}                                            \\
        \hline
        \textbf{Project} & \textbf{Instruction} & \textbf{Line}   & \textbf{Branch} & \textbf{Method} \\
        \hline
        math             & 6,007.28
 (486.28)      & 376.24

 (42.28) & 140.41
 (13.14)  & 55.45
 (7.86)      \\
        joda-time        & 10,139.60
 (22.40)          & 2,455.20
 (4.40)    & 771.40

 (3.80)    & 568.80

 (0)     \\
        jfreechart       & 449.50

 (61)             & 122

 (16)       & 21.50

 (4.50)     & 33.50

 (5)        \\
        jackson          & 3,970

(9)
             & 856.50

 (2.50)        & 279.50

 (3)        & 140
 (0)       \\
        fastjson         & 10,705.25(91.25)            & 1,827

 (12)        & 616

(11.25)         & 221.50

 (1.25)         \\
        lang             & 474.33

 (6.55)             & 67
 (1.72)   & 26.06 (1.62)
   & 12.17
 (0.11)
 \\
        \hline
Avg Imp & 4.5\%&	5.3\% &	7.8\% &	5.9\% \\
\hline

        \end{tabular}
    }
    \begin{tablenotes}\footnotesize
        \item* Note that the results of \slicing and \trycatch on coverage are the same. The value in ``()'' indicates the extra coverage achieved by \trycatch and \slicing over \original. Avg Imp stands for average improvement over different projects.
    \end{tablenotes}
   
    \end{threeparttable}
     \vspace{-0.2in}
\end{table}

\noindent\textbf{\trycatch and \slicing improve the line/branch coverage in 55\% (30/60) of the studied versions compared to the \original. }
To provide a better view of how many versions get improved, we compare the coverage of \original, \trycatch, and \slicing for every version. Table~\ref{tab:RQ2_versionwise} presents the number of versions in which the coverage gets improved, deteriorated, and tied after applying \trycatch and \slicing compared with \original in terms of four granularities, i.e., instruction, line, branch, and method. First of all, the coverage of no version gets deteriorated in all granularities, which is expected, since \trycatch and \slicing enforces \testearlystop to keep executing even if an assertion failure happens. In more than half of the versions, the instruction (34/60), line (33/60), and branch (33/60) coverage gets improved. In 8 out of 60 versions, the method coverage also gets improved.

In summary, our findings indicate that different assertions in a test verify different parts of the program, \earlystop would miss the verification of certain production code. 

\begin{table}[]
    \caption{The number of versions that get improved and tied after applying \trycatch compared with \original in terms of different coverage granularities (i.e., Instruction (Ins), Line (Ln), Branch (Br), and Method (Md)).}
    \label{tab:RQ2_versionwise}
     \centering
    \resizebox{0.4\textwidth}{!}{

    \begin{tabular}{l|rrrr|rrrr}
    \hline
    \textbf{Project}   & \multicolumn{4}{c}{\textbf{\#Improved}}    &  \multicolumn{4}{c}{\textbf{\#Tied}}      \\
    \hline
              & \textbf{Ins} & \textbf{Ln} & \textbf{Br} & \textbf{Md} & \textbf{Ins} & \textbf{Ln} & \textbf{Br} & \textbf{Md} \\
    \hline
    math       & 17 & 18 & 16 & 6 & 12 & 11 & 13 & 23 \\
    joda-time  & 2  & 2  & 4  & 1 & 3  & 3  & 1  & 4  \\
    jfreechart & 1  & 1  & 0  & 1 & 1  & 1  & 2  & 1  \\
    jackson    & 2  & 2  & 1  & 0 & 0  & 0  & 1  & 2  \\
    fastjson   & 4  & 4  & 4  & 0 & 0  & 0  & 0  & 4  \\
    lang       & 8  & 6  & 8  & 0 & 10 & 12 & 10 & 18 \\
    \hline
    \textbf{Total}      & 34 & 33 & 33 & 8 & 26 & 27 & 27 & 52  \\
    \hline
    \end{tabular}
    }
    \begin{tablenotes}\footnotesize
    \centering
        \item* Note that the number of versions that get deteriorated is zero across all studied projects. To save space, we do not show the results of deteriorated cases.
    \end{tablenotes}
\vspace{-0.1in}
\end{table}

\rqboxc{In general, \earlystop negatively impacts the coverage in the production code in terms of all granularities (i.e., instruction, line, branch, and method). \trycatch and \slicing improve the line/branch coverage in 55\% (33/60) of the studied versions compared with \original.}

\subsection*{\rqthree}\label{sec:rq3}

\begin{table*}
\footnotesize
\caption{Comparison of the performance of SBFL in different settings: \original, \trycatch, and \slicing in terms of MFR, EXAM, Top@k (k = 5 and 10) for {\bf Ochiai}. The columns presenting the best result are marked in bold.}\label{tab:FLresults}
\resizebox{0.95\textwidth}{!}{
\begin{tabular}{l|rrrr|rrrr|rrrr}
\hline
\multicolumn{5}{c}{\textbf{Original}}                                                  & \multicolumn{4}{c}{\textbf{Trycatch}}                                & \multicolumn{4}{c}{\textbf{Slicing}}                                \\
\hline
\multirow{2}{*}{\textbf{Project}} & \multirow{2}{*}{\textbf{MFR}} & \multirow{2}{*}{\textbf{EXAM}} & \multicolumn{2}{c|}{\textbf{Top@k}} & \multirow{2}{*}{\textbf{MFR}} & \multirow{2}{*}{\textbf{EXAM}} & \multicolumn{2}{c|}{\textbf{Top@k}} & \multirow{2}{*}{\textbf{MFR}} & \multirow{2}{*}{\textbf{EXAM}} & \multicolumn{2}{c}{\textbf{Top@k}} \\
                 &                &                & \textbf{5}   & \textbf{10}   &                &                & \textbf{5}   & \textbf{10}   &                &                & \textbf{5}   & \textbf{10}   \\
\hline
math     & 15.2  & 0.00191 & 25.9\% & 30.1\% & 11.7  & 0.00188    & \textbf{30.9\%} & \textbf{34.3\%} & \textbf{9.1}  & \textbf{0.00183} & 26.0\%      & 33.0\%      \\
joda-time  & 596.2 & 0.03306 & 0.0\% & 0.0\%  & 562.6 & 0.03318    & 0.0\%      & 0.0\%      & \textbf{431.4} & \textbf{0.02733} & 0.0\%      & 0.0\%      \\
jfreechart  & 125.5 & 0.00136 & 0.0\% & 0.0\%  & 125.5 & 0.00136    & 0.0\%      & 0.0\%      & 125.5      & 0.00136     & 0.0\%      & 0.0\%      \\
jackson-core & 633.0 & 0.0383 & 0.0\% & 0.0\%  & 631.5 & 0.03821    & 0.0\%      & 0.0\%      & \textbf{627.5} & \textbf{0.03796} & 0.0\%      & 0.0\%      \\
fastjson   & 7,198.3 & 0.16767 & 0.0\% & 0.0\%  & 7,182.5 & 0.167 & 0.0\%      & 0.0\%      & \textbf{7,175.3} & 0.167  & 0.0\%      & 0.0\%      \\
lang     & 30.7  & 0.00237 & 6.9\% & 10.23\% & 27.6  & 0.00236    & 6.9\%     & 10.2\%     & \textbf{23.9}  & \textbf{0.00228} & \textbf{10.7\%} & \textbf{14.90\%} \\
\hline
\end{tabular}
}
\end{table*}

\begin{table*}
\footnotesize
\caption{Comparison of the performance of SBFL in different settings: \original, \trycatch, and \slicing in terms of MFR, EXAM, Top@k (k = 5 and 10) for \textbf{Tarantula}. The columns presenting the best result are marked in bold.}\label{tab:FLresultsTarantula}
\resizebox{0.95\textwidth}{!}{
\centering
\begin{tabular}{l|rrrr|rrrr|rrrr}
\hline
\multicolumn{5}{c}{\textbf{Original}}                                                  & \multicolumn{4}{c}{\textbf{Trycatch}}                                & \multicolumn{4}{c}{\textbf{Slicing}}                                \\
\hline
\multirow{2}{*}{\textbf{Project}} & \multirow{2}{*}{\textbf{MFR}} & \multirow{2}{*}{\textbf{EXAM}} & \multicolumn{2}{c|}{\textbf{Top@k}} & \multirow{2}{*}{\textbf{MFR}} & \multirow{2}{*}{\textbf{EXAM}} & \multicolumn{2}{c|}{\textbf{Top@k}} & \multirow{2}{*}{\textbf{MFR}} & \multirow{2}{*}{\textbf{EXAM}} & \multicolumn{2}{c}{\textbf{Top@k}} \\
                 &                &                & \textbf{5}   & \textbf{10}   &                &                & \textbf{5}   & \textbf{10}   &                &                & \textbf{5}   & \textbf{10}   \\
\hline
math     & 16.4  & 0.00196
 & 25.9\% & 29.6\% & 13.1
  & 0.00190    & \textbf{32.6\%} & \textbf{36.1\%} & \textbf{10.6}  & \textbf{0.00185} & 32.4\%      & 36.1\%      \\
joda-time  & 586.8
 & 0.03051 & 0.0\% & 0.0\%  & 583.2
 & 0.03002    & 0.0\%      & 0.0\%      & \textbf{469.6} & \textbf{0.02715} & 0.0\%      & 0.0\%      \\
jfreechart  & 125.5
 & 0.00135 & 0.0\% & 0.0\%  & 125.5 & 0.00135    & 0.0\%      & 0.0\%      & 125.5      & 0.00135     & 0.0\%      & 0.0\%      \\
jackson-core & 633.0
 & 0.0383 & 0.0\% & 0.0\%  & 627.5
 & 0.03796    & 0.0\%      & 0.0\%      & \textbf{627.5} & \textbf{0.03796} & 0.0\%      & 0.0\%      \\
fastjson   & 7,271.0
 & 0.16937 & 0.0\% & 0.0\%  & 7232.3
 & 0.16846 & 0.0\%      & 0.0\%      & \textbf{7,221.5} & \textbf{0.16821}  & 0.0\%      & 0.0\%      \\
lang     & 25.5
  & 0.00226 & 10.7\% & 25.3\% & 25.3  & 0.00225    & 9.3\%     & 24.0\%     & \textbf{23.7}  & \textbf{0.00213} & \textbf{10.90\%} & \textbf{29.2\%} \\
\hline
\end{tabular}
}
\end{table*}


\textbf{In general, \earlystop negatively impacts the effectiveness of SBFL. Eliminating early test termination improves the performance of Ochiai and Tarantula in all evaluation metrics. For instance, Both \slicing and \trycatch can improve the performance of SBFL by 15.1\% and 6.7\% on average in terms of MFR, respectively, compared with \original for Ochiai.} Tables~\ref{tab:FLresults} and \ref{tab:FLresultsTarantula} present the performance of SBFL in \original, \trycatch, and \slicing in terms of MFR, EXAM, and Top@k for Ochiai and Tarantula, respectively. In general, we observe that both \slicing and \trycatch outperform \original in terms of MFR and EXAM in all projects for both Ochiai and Tarantula, except jfreechart. In other words, \earlystop negatively impacts the effectiveness of SBFL. Typically, on average, \slicing improves the performance of Ochiai and Tarantula by 15.1\% and 10.7\% compared with \original in terms of MFR across all projects, respectively. Particularly, in the two largest projects math and lang, MFR is reduced by \earlystop for both SBFLs (Ochiai and Tarantula) by a large margin. In math, MFR is deteriorated from 9.7 (\slicing) to 15.2 by 40.1\% for Ochiai and is reduced from 10.6 to 16.4 by 35.4\% for Tarantula, respectively. 
In terms of Top@k, \slicing and \trycatch improve \original in both math and lang. We do not see improvement in other projects in terms of Top@k. 
Although MFR has improved, none of the fault statements could be identified in the top 10 results. 
Similar to RQ2, as shown in Table~\ref{tab:FL_versionwise}, we also observe that in a remarkable portion of the versions, \trycatch and \slicing improve the effectiveness of SBFL compared with \original, for both Ochiai and Tarantula, respectively. For instance, \slicing outperforms \original in 18 and 22 out of 60 versions for Ochiai and Tarantula, respectively. 

Let's consider an example that illustrates why eliminating \earlystop (i.e., \trycatch) could potentially improve the effectiveness of SBFL in some cases. In version 72 of math, the SBFL gets improved after applying \trycatch. The test shown in Figure~\ref{lst:example1} aims to test the function $solve()$ shown in Figure~\ref{lst:successfulcase}, which has two buggy lines located in two branches. In \original, the test terminates at the third assertion (line 13) in the test and misses the opportunity to reveal the bug (line 16) in the second branch (lines 15 - 18) in function $solve()$, which is verified by the last assertion in the test (line 17) due to \earlystop. While in \trycatch, all assertions are executed and both of the buggy lines are identified by SBFL.


\begin{figure}
  \begin{lstlisting}[language=Java]
public double solve(final UnivariateRealFunction f, final double min, final double max, final double initial) throws MaxIterationsExceededException, FunctionEvaluationException {
     ...
    // return the first endpoint if it is good enough
    double yMin = f.value(min);
    if (Math.abs(yMin) <= functionValueAccuracy) {
      setResult(yMin, 0);  //buggy line
      return result;
    }
    // reduce interval if min and initial bracket the root
    if (yInitial * yMin < 0) {
      return solve(f, min, yMin, initial, yInitial, min, yMin);
    }
    // return the second endpoint if it is good enough
    double yMax = f.value(max);
    if (Math.abs(yMax) <= functionValueAccuracy) {
      setResult(yMax, 0); //buggy line
      return result;
    }
    ...
} \end{lstlisting}
  \caption{The function which is tested by the code in Figure~\ref{lst:example1}.}
  \label{lst:successfulcase}
\end{figure}

\textbf{\slicing outperforms or has the same performance as \trycatch for SBFL (both Ochiai and Tarantula) in terms of MFR and EXAM in all projects. For Ochiai, \slicing improves \trycatch by 9.8\% and 4.1\% in terms of MFR and EXAM, respectively.}
\slicing outperforms or has the same performance as \trycatch in terms of EXAM and MFR in all studied projects as shown in Table~\ref{tab:FLresults}. On average, for Ochiai, \slicing improves \trycatch by 9.8\% and 4.1\% in terms of MFR and EXAM, respectively. Let's consider the two largest projects math and lang. In math, \slicing improves MFR and EXAM scores both by 21.6\% and 2.6\% for Ochiai compared with \trycatch, respectively. In lang, \slicing improves MFR and EXAM by 13.1\% and 3.4\% for Ochiai, respectively. In terms of Top@k, for Ochiai, in lang, we observe that \slicing outperforms \trycatch by 54.9\% and 45.7\% when k equals 5 and 10, respectively. In other relatively smaller projects, \slicing has better or equivalent performance as \trycatch. We observe a similar trend for Tarantula. If we compare \slicing with \original, the improvement is even more remarkable. For instance, on average, for Ochiai, \slicing improves the \original by 15.1\% and 4.4\% in terms of MFR and EXAM, respectively. \slicing improves the performance of SBFL in 28.3\% and 30\% of studied versions over \trycatch for Ochiai and Tarantula, respectively. We present the comparison between \slicing and \trycatch in  Table~\ref{tab:FL_versionwise}. We observe that \slicing outperforms \trycatch in 17 and 18 out of 60 versions for Ochiai and Tarantula, respectively. Overall, \slicing outperforms \trycatch in terms of both SBFLs.



\begin{table}[]
\footnotesize
\caption{The number of versions that get improved (Imp), deteriorated (Det), and tied after applying Trycatch and Slicing compared with original for Ochiai (first value in each cell) and Tarantula (second value) in terms
of MFR, respectively.}
\label{tab:FL_versionwise}

\resizebox{0.5\textwidth}{!}{
\begin{tabular}{l|rrr|rrr|rrr}
\hline
     & \multicolumn{3}{c|}{\textbf{Original vs Trycatch}}    & \multicolumn{3}{c|}{\textbf{Original vs Slicing}}    & \multicolumn{3}{c}{\textbf{Trycatch vs Slicing}}    \\
\hline
\textbf{Project} & \textbf{\#Tied} & \textbf{\#Imp} & \textbf{\#Det} & \textbf{\#Tied} & \textbf{\#Imp} & \textbf{\#Det} & \textbf{\#Tied} & \textbf{\#Imp}  &\textbf{\#Det} \\
\hline
math     & 23/23 & 3/3 & 3/3 & 23/23 & 5/5 & 1/1 & 22/23 & 4/4 & 3/2 \\
joda-time  & 3/3 & 1/1 & 1/1 &  0/0  & 5/5 & 0/0 & 0/0 & 5/5 & 0/0 \\
jfreechart  & 1/2 & 0/0 & 1/0 & 1/2  & 0/0 & 1/0 & 2/2 & 0/0 & 0/0 \\
jackson-core & 1/1 & 1/1 & 0/0 &  1/1  & 1/1 & 0/0 & 1/2 & 1/0 & 0/0 \\
fastjson   & 0/0 & 3/4 & 1/0 &  0/0  & 3/4 & 1/0  &0/1 & 1/2 & 3/1 \\
lang     & 12/12 & 3/3 & 3/3 &   12/8 & 4/7 & 2/3 & 10/8 & 6/7 & 2/3 \\
\hline
\textbf{Total}    & 35/36 & 11/12 & 9/7 & 37/34 & 18/22 & 5/4 & 37/34 & 17/18 & 8/6 \\
\hline
\end{tabular}
}

\vspace{-0.2in}
\end{table}

Now let us use an example to illustrate why \slicing outperforms \trycatch. For instance, in version 41 of the project lang, there are 1,624 runnable tests after applying \trycatch. Among those tests, there are two failed tests. However, after applying \slicing, the number of total runnable tests is increased to 1661, with a total of 17 failures among them, which indicates there are 15 more failed tests. In this case, the buggy lines are only executed in two failed tests when applying \trycatch, while they are executed in 17 failed tests when applying \slicing. When the number of those failed tests is plugged into the Ochiai and Tarantula formulas, we get the rank of the first buggy line for \trycatch and \slicing at 7 and 3 for Ochiai, and at 8 and 5 for Tarantula. \slicing alleviates the limitation of \trycatch (i.e., aggregating all assertion failures in a test into a single testing result) and provides a finer-grained test result.

\rqboxc{In general, \earlystop negatively impacts the effectiveness of SBFL. Both \slicing and \trycatch improve the performance of SBFL. \slicing outperforms \trycatch for SBFL. On average, \slicing improves the performance of Ochiai and Tarantula by 15.1\% and 10.7\% compared with \original in terms of MFR, respectively. } 

\section{Discussion}\label{sec:discussion}
\subsection{Overhead of our approaches}

The overhead of Slicing and TryCatch is low as shown in Table~\ref{tab:overhead} (on average, two seconds for each system). We only applied test transformation on failed test cases with early termination and rerun them, which usually takes only milliseconds. Therefore, the overhead mainly comes from test execution. Slicing introduces extra overhead since it splits a test case with n assertions into n sub-test cases. TryCatch adds try-catch blocks to force the execution of the entire test case. In theory, TryCatch brings less overhead than Slicing, since Slicing increases the size of test cases, while TryCatch does not.

\begin{table}[]
\centering
\scriptsize
\caption{The average running time (in seconds) of the test suite for each project across versions in three settings \original, \trycatch, and \slicing.}\label{tab:overhead}
\begin{tabular}{l|r|r|r}
\hline
\textbf{Project}   & \textbf{Original} &\textbf{TryCatch} & \textbf{Slicing} \\
\hline
\textbf{math}      & 12.96                               & 13.16                               & 13.29                              \\
\textbf{joda-time} & 7.02                                 & 7.14                                 & 7.23                                \\
\textbf{jfree}     & 5.46                                 & 5.47                               & 5.49                               \\
\textbf{jackson}   & 4.36                                & 4.41                                & 4.44                                \\
\textbf{fastjson}  & 43.83                                & 56.62                                & 56.94                             \\
\textbf{lang}      & 12.50                               & 12.56                               & 12.67                             \\
\hline
\end{tabular}
\vspace{-0.2in}
\end{table}

\subsection{Implications of our findings}

\noindent\textbf{We recommend that developers limit each test to a single assertion. In cases where this is not feasible, use \slicing to eliminate \earlystop.} In the recent versions of JUnit, certain mechanisms have been developed, such as \textit{ErrorCollector} starting from JUnit 4 and \textit{assertAll} starting from JUnit 5, to enable the test to keep executing even if an assertion fails in the middle. Similar to \trycatch, although such mechanisms (\textit{ErrorCollector} and \textit{assertAll}) can avoid \earlystop, our findings show that \slicing provides more accurate testing information than other mechanisms and outperforms \trycatch in locating faults. This is because \slicing disassembles a test with multiple assertions into a set of independent sub-tests, that enable the execution of every assertion and its associated code independently and provide fine-grained granularity. Therefore, we strongly encourage developers to limit each test to a single assertion whenever possible and to apply \slicing wherever multiple assertions are necessary. Although, one limitation of \slicing is that it splits one test into multiple tests and code before assertions are executed multiple times which introduces additional overhead.

\textbf{We encourage future research to investigate how the industry solutions (e.g., \textit{assertAll} and \textit{ErrorCollector}) are applied for handling multiple assertions in a test case.} Although having multiple assertions in a single test can be considered a test smell and not recommended, this is a common practice in the studied projects. It is also interesting to understand why developers do (not) use the \textit{assertAll} (Junit 5) and \textit{ErrorCollector} (Junit 4) mechanisms and facilitate the use of such mechanisms.

\subsection{Threats to Validity}\label{sec:threats}
\noindent\textbf{Internal Validity}
In RQ3, we evaluate SBFL using three commonly used metrics (i.e., EXAM, top@k, and MFR). Although there might be other metrics that could be used for our task, these metrics are commonly used in evaluating SBFL techniques~\cite{widyasari2022real,pearson2017evaluating}. A threat to our study is that the program transformation possibly change the behavior of test cases. To mitigate this threat, we measure the quality of transformed tests, we applied \trycatch and \slicing on passed test cases with multiple assertions and measured their code coverage. Our results show that \original, \trycatch, and \slicing have almost the same code coverage (at most 0.4 lines difference across all tests), which means the behaviour of the tests remains mostly the same. In addition, our transformation does not change the results from passed to failure. See more detailed results in our replication package~\cite{DataRepository}.


\noindent\textbf{External Validity}
Threats to external validity relate to the generalizability of our findings.
In RQ3, we selected two popular SBFL techniques Ochiai and Tarantula as our subject techniques. {\it Early test termination} may have a different influence on the effectiveness of other SBFL techniques. However, our proposed framework could be applied to any techniques that rely on test coverage information. Another threat relates to our studied projects. Our findings might not be generalized to other projects. To mitigate the threat, we selected 6 projects that are in various domains from popular benchmarks Defect4J and T-Evos. Among all the 471 versions, 60 versions contain 90 \testearlystop. Future research may further expand the study on more datasets. 
In this study, we focus on early test termination due to assertion failure (only 19\% of the studied early test termination), there are other reasons (e.g., error in method invitation and object construction) that could lead to early test termination. We encourage future research to investigate other reasons.

\section{Related work}\label{sec:relatedwork}

\subsection{Understanding and improving the usage of assertion in software testing.}
Xuan et al. proposed an approach to separate existing test cases into small fractions (called purified test cases) for each assertion, which is similar to \slicing, and to enhance the test oracles to further localize faults~\cite{xuan2014test}. This work is the most closely related to our study. However, different from Xuan et al's study which focuses on improving the effectiveness of fault localization by purifying test cases, our study focuses on investigating the impact of \earlystop. More specifically, we investigated the prevalence of \earlystop and provided empirical evidence for it. Moreover, we investigated the impact of \earlystop on code coverage and fault localization by comparing it with the baseline in which we eliminated \earlystop using \trycatch.

A few studies have been conducted to understand assert practice in testing~\cite{bai2022assertion,zamprogno2022dynamic,jia2021unit,bavota2012empirical}. Bavota et al. studied 18 systems to analyze the distribution of test smells and show that Assertion Roulette (i.e., a code smell occurs when a test method has multiple non-documented assertions) is presented in 62\% of the total JUnit classes~\cite{bavota2012empirical}. Zamprogno et al. examined more than 30k assertions from 105 projects and also found 17.4\% of the test cases have three or more assertions, which is compatible with our finding in RQ1. 
On the other hand, various solutions were proposed to improve the assert practice in unit testing~\cite{watson2020learning,tufano2022generating,yu2022automated,song2007unitplus,fraser2011evosuite}. For instance, Fraser and Arcuri proposed EvoSuite, which uses mutation testing to produce a set of assertions that maximizes the number of seeded defects in a class that can be revealed by the test cases. Watson et al. employed a Neural Machine Translation (NMT) based approach called Atlas to automatically generate meaningful assert statements providing a test method and a focal method to test the correctness of the focal method~\cite{watson2020learning}.
Our study differs from previous ones in that, we concentrate on examining a specific occurrence of test failure, which arises when an assert statement fails in the middle of a test case, thereby preventing the execution of code after that point. 

\subsection{Investigating the factors that influence testing coverage and the corresponding techniques}
Masri et al. investigated the prevalence of factors that impair the effectiveness of coverage-based fault localization~\cite{masri2009empirical}. They found that 72\% of their studied buggy versions exhibit the condition for failure was met but the program did not fail. More specifically, three conditions must be satisfied for program failure to occur: (1) the defect's location must be reached, (2) the program's state must become infected and (3) the infection must propagate to the output.
Several studies have been done on this and demonstrate the prevalence of Coincidental correctness and its negative impact on fault detection and localization techniques~\cite{hierons2006avoiding,wang2009taming,abou2021detrimental,masri2009empirical}. 
For instance, Assi et al. investigated the impact of coincidental correctness on test suite reduction (TSR), test case prioritization (TCP), and SBFL~\cite{abou2021detrimental}. They found that TSR and TCP, the negative impact of CC was very significant. 

Different from prior studies, we focus on investigating the impact of \earlystop on the coverage and corresponding techniques that rely on it (i.e., SBFL). Our findings show that \earlystop does negatively impact code coverage and the effectiveness of SBFL.

\section{Conclusion}
\label{sec:conclusion}
In this paper, we first study the prevalence of \earlystop in failed tests and conduct experiments on six widely used open-source projects in software debugging and testing research from Defects4J and T-Evos benchmarks. Our findings show that 50.3\% of failed tests are early terminated, with 19.1\% of them caused by assertion failure.
We also demonstrate that \earlystop negatively impacts the code coverage across all granularities and can affect the performance of SBFL.
To address this issue, we propose to use two approaches, \trycatch and \slicing, that catch assertion failures and enforce test execution to mitigate the impact of early test termination.
Our evaluation indicates that \slicing and \trycatch can improve the performance of SBFL by 15.1\%, 6.7\%  for Ochiai, and 10.66\%, 3.81\% for Tarantula on average in terms of MFR, respectively.
Our study emphasizes the impact of \earlystop on code coverage and the performance of fault localization techniques that rely on it. We also suggest that practitioners limit each test to a single assertion or use \slicing when multiple assertions are necessary.







\section{Data Availability}
We made our dataset publicly available~\cite{DataRepository} to encourage future research on studying the impact of test quality on fault localization and related approaches.

\bibliographystyle{IEEEtran}
\scriptsize{
\bibliography{main}
}

\end{document}